\documentclass[12pt,a4]{iopart}
\usepackage{iopams}  

\usepackage{epsfig,rotating}
\usepackage{graphicx}
\usepackage{epstopdf,color}

\def\cm{cm$^{-1}$\,}
\newcommand{\bra}[1]{\left<#1\right|}
\newcommand{\ket}[1]{\left|#1\right>}

\newcommand{\Eq}[1]{Eq.~(\ref{#1})}
\providecommand\ket[1]{| #1 \rangle}\providecommand\bra[1]{\langle #1 |}

\begin{document}

\title{Exciton coupling induces vibronic hyperchromism in light-harvesting complexes}

\author{Jan Schulze$^1$, Magne Torbj\"ornsson$^2$, Oliver K\"uhn$^1$, T\~onu Pullerits$^2$}
\address{$^1$Institut f\"{u}r Physik, Universit\"{a}t Rostock, D-18051 Rostock, Germany}
\address{$^2$Department of Chemical Physics, Lund University, P.O. Box 124, 22100 Lund, Sweden}

\ead{tonu.pullerits@chemphys.lu.se, oliver.kuehn@uni-rostock.de}

\begin{abstract}
  The recently suggested possibility that weak vibronic transitions can be excitonically enhanced in
  light-harvesting complexes is studied in detail. A vibronic exciton dimer model which includes ground state
  vibrations is investigated using multi-configuration time-dependent Hartree method with a parameter set
  typical to photosynthetic light-harvesting complexes. Absorption spectra are discussed in dependence on the
  Coulomb coupling, the detuning of site energies, and the number of vibrational mode. Calculations of the
  fluorescence spectra show that the spectral densities obtained from the low temperature fluorescence line
  narrowing measurements of light-harvesting systems need to be corrected for the exciton effects. For the
  J-aggregate configuration, as in most of the light-harvesting complexes, the true spectral density has
  larger amplitude than what is obtained from the measurement.
\end{abstract}

\maketitle

\section{Introduction}

In photosynthesis light absorption and charge separation take place in specialised multi-chromophoric
systems. While the reaction centre (RC) complexes, where the primary charge separation occurs, are very
similar in different photosynthetic systems, the light-harvesting antennas show a wide variety
\cite{BlankenshipBook}. Clearly, it is possible to optimise absorption of light and make excitation transport
to RC efficient in many different ways. One of the strategies used in most of the photosynthetic systems is to
organise the pigments energetically in an energy funnel with the RC in the bottom of the sink
\cite{Pullerits1996}. In addition, the pigment molecules are usually close enough to enable sufficient
resonant interaction and thereby fast excitation transport. Such general optimisation strategies are widely
reported and well understood in photosynthetic antenna systems.

A result of the relatively dense packing of the pigment molecules in the antenna complexes is the emergence of
delocalised Frenkel exciton states~\cite{MayKuhnBook,ExcitonBook,renger01_137,Kuhn2002}. The later has led to
recognition of the role of memory effects and coherence in Frenkel exciton dynamics in photosynthetic
light-harvesting. The discussion was recently revitalised by novel coherent multidimensional spectroscopy
measurements revealing long-lived oscillatory features in a so called Fenna-Mathews-Olson (FMO)
light-harvesting antenna complex \cite{Engel2007}. The oscillations were interpreted as electronic coherences
and their long lifetime was taken as an evidence for possible coherent excitation transport. It was argued
that parallel coherent excitation transport pathways may enable interference-based quantum optimisation
possibilities for making excitation transfer through the FMO complexes more efficient. Analogous observations
were also reported for light-harvesting complexes in photosynthetic marine algae \cite{Collini2010}. This
initiated numerous new theoretical investigations of the role of coherence in excitation energy transport
\cite{Fassioli2010, Ishizaki2009, Abramavicius2009}. An important conclusion drawn from these efforts
concerned the optimal parameter region of the transport - it was realized that the most efficient transport
takes place in the intermediate regime between fully coherent and incoherent transport where the inter-pigment
resonance interaction and the system-bath interaction are not so different \cite{Plenio,
  Aspuru}. Interestingly, the efficiency curve has a rather flat maximum meaning that optimal transport can be
achieved by quite a broad set of parameters.

In contrast to the large number of simulations following the excitation dynamics \cite{Ishizaki2009,
  ShimAspuru, Coker, Ishizaki2009a, Nalbach2011}, much fewer studies have made the effort to calculate the coherent multidimensional
spectroscopy signals, which would directly correspond to what is observed in experiment \cite{BenCPL07,
  kjellberg, cheng}. It turned out that reproducing the long-lived oscillations in calculated 2D spectra as electronic coherences is
not straightforward. Even studies based on hierarchy equations of motion (HEOM) where bath memory effects
are included formally exact could not reach agreement with experimental observations \cite{Chen,
  kramer}. Additional assumptions had to be made. For example, assuming correlated static and dynamic disorder
at different pigment sites led to long-lived electronic coherences \cite{Nalbach2010, Abramavicius2011}.  However, neither
correlated nuclear motions nor inhomogeneous broadening has any independent support. Quite the
opposite. Thorough molecular dynamics studies did not show any evidence for such correlations \cite{Ulrich}. A
recent study using HEOM approach produced long-lasting electronic coherences \cite{Kramer12}. The authors used
a spectral density, which fitted well the fluorescence line narrowing experiments \cite{Wendling2000} for
frequencies above 50 cm$^{-1}$. However, the low-frequency region, which is particularly important for
electronic dephasing and is most sensitive to temperature, had an unrealistically small amplitude.

Initially, the possible vibrational origin of the oscillatory features in the two-dimensional (2D) signal in
FMO was discarded since the nuclear modes in the frequency region of observed oscillations ($\sim$150
cm$^{-1}$) are very weakly coupled to the electronic transition leading to a negligible Huang-Rhys factor
$S$. In the 180 cm$^{-1}$ region there are a few modes with the total $S$ reaching almost 0.05
\cite{Wendling2000, Freiberg2011}, but even this is too weak to produce significant oscillatory features in 2D
signals. Transition energies of the bacteriochlorophyll (BChl) molecules in FMO are quite well established
\cite{Adolphs_Renger2006}. It turns out that the 0-1 vibronic transition energy of the 180 cm$^{-1}$ mode
region on the red-most BChl is almost in resonance with the 0-0 transition energy of one of its
neighbours. Consequences of such resonance for natural light-harvesting have not been fully analysed before. A
vibronic exciton model where one mode per pigment is treated explicitly \cite{Sergey} was applied for
calculating electronic 2D response functions of FMO \cite{Niklas_FMO}. It was shown that owing to the
resonance, the weak transition of mainly vibronic origin at the lowest energy BChl obtains a significant
additional oscillator strength due to mixing with the strong purely electronic transition at the neighbouring
molecule. Furthermore, the long-lived oscillations observed in 2D experiments were explained by the fact that
energy fluctuations of the vibrational levels of a molecule are correlated and consequently the corresponding
coherences between vibronic levels have a long lifetime. The intensity borrowing is quite robust and moving to
some extent out from the resonance due to the inhomogeneous broadening does not influence the results
appreciably.

An analogous resonance between 0-0 and 0-1 transitions of neighbouring pigments was recently used in an
excitonic dimer model where adiabatic potential energy surfaces were calculated
\cite{Tiwari2013,PNAS2013}. The authors argue that because of the near resonance between the levels, the
adiabatic approximation breaks down and non-adiabatic coupling has to be taken into account. The non-adiabatic
coupling causes mixing of the states leading to intensity redistribution very much like in
Ref. \cite{Niklas_FMO}. Excitation dynamics due to the non-adiabatic coupling in the context of
light-harvesting complexes has been analysed in terms of curve-crossing and surface hopping \cite{Beenken2002,
  Dahlbom2002}. Schr\"oter and K\"uhn studied the interplay between non-adiabatic dynamics and Frenkel exciton
transfer in aggregates where both S$_1$ and S$_2$ transitions were considered \cite{Schroter2013}. Also
excitation annihilation has been described as non-adiabatic coupling between one- and two-exciton manifolds
\cite{an0, Bruggemann2009}. Tiwari et al.  \cite{Tiwari2013} investigated a different situation where the
potentials are nested and the crossing points are very far from the equilibrium resembling the situation in
internal conversion. Using 2D signal calculations the authors argue that the long-lived oscillating features
of 2D spectroscopy of light-harvesting systems could be explained in terms of ground state coherent nuclear
motions excited via enhanced transitions with strong vibronic character.

Several other studies have recently addressed the issue how to separate the electronic and vibrational quantum
beats in electronic 2D spectroscopy. Comparison of the oscillatory patterns and oscillation phases in two
representative systems, displaced oscillator and excitonic dimer, was reported \cite{Butkus2012}. The model
was further developed for a dimer with ground state vibrational levels included \cite{Butkus2013}. The ground
and excited state vibrational coherences were thoroughly compared in vibronic exciton dimer model
\cite{Chenu2013}. Oscillatory phonon features in colloidal quantum dots were modelled \cite{Seibt2013} and it
was shown that following separately positive and negative frequency components of the population time Fourier
transform gives additional control over the Liouville pathways thereby enabling a more clear distinction
between electronic and vibrational beatings \cite{Seibt2013a}.

We point out that the method used by Christensson et al. \cite{Niklas_FMO}, a so called one particle
approximation (OPA), only includes excited state vibrations. In order to account for the ground state
vibrational states, at least the two particle approximation (TPA) \cite{Philpot, Spano} has to be
employed. TPA is exact in case of a dimer and has been used for analyses of 2D spectra \cite{Tiwari2013,
  Butkus2013}. Various analogous approaches have been applied for modelling a vibronic dimer \cite{Fulton,
  Eisfeld2005}. An obvious question arrises - what effects does one miss by using simple OPA calculations as
in \cite{Niklas_FMO} compared to the exact solution of the problem. Here we address the issue by carrying out
a comparison of OPA and TPA using a model dimer resembling two neighbouring low-energy BChl molecules in the
FMO. The numerically exact reference is provided by the multi-configuration time-dependent Hartree (MCTDH)
method \cite{meyer90:73,beck00:1}.

The article is organised as follows. We start from presenting the basics of the the vibronic exciton model and
the OPA as well as MCTDH approaches. The theory is followed by comprehensive model calculations of a vibronic
heterodimer for various parameter sets. We show that despite of the dominantly monomeric character of the
heterodimer, the intensities of the vibrational features in the fluorescence spectrum are significantly
affected. This has consequences for how to use the fluorescence line narrowing spectroscopy to experimentally
determine the spectral densities \cite{Pullerits1994}. In the final part the results are discussed and
conclusions formulated.
\section{Theoretical Model}\label{sec:main} 
\subsection{Frenkel Exciton Hamiltonian}
\label{sec:model-hamiltonian}
In the following we will use the Frenkel exciton Hamiltonian describing coupled electronic (zero, $|0\rangle$,
and one-exciton, $|m\rangle$, space) and nuclear degrees of freedom, $\mathbf{Q}=\{\mathbf{Q}_m\}$,
\cite{MayKuhnBook}
\begin{eqnarray}
  H_{\rm agg}(\mathbf{Q}) &=& H^{(0)}(\mathbf{Q}) + H^{(1)}(\mathbf{Q}) \, , \\
  H^{(0)}(\mathbf{Q}) &=&  \sum_m H_{m,g}(\mathbf{Q}_m) \ket{0}\bra{0} \equiv \mathcal{E}_0\ket{0}\bra{0} \, ,\\
  H^{(1)} (\mathbf{Q})& = &  \sum_{mn}[ \delta_{mn}(\mathcal{E}_0 + U_{m,e}(\mathbf{Q}_m) +  J_{mn}] |m\rangle \langle n| \, .
\end{eqnarray}
Denoting the monomeric electronic states by $\ket{a_m}$ where $a=(g,e)$ are the ground and excited states, one
has $\ket{0}= \prod_m \ket{g_m}$ and $\ket{m}= \ket{e_m}\prod_{n \ne m} \ket{g_n}$.  The nuclear motion will
be describe in the displaced oscillator model, i.e. the ground state Hamiltonian and excited state coupling
read for each site ($E_{m}$: electronic energy)
\begin{eqnarray}
  H_{m,g}(\mathbf{Q}_m) &=& \sum_\xi\frac{\hbar\omega_\xi}{2} \left( - \frac{\partial^2}{\partial Q_{m,\xi}^2}+ Q_{m,\xi}^2\right) \, ,\\
  U_{m,e}(\mathbf{Q}_m) & = & E_{m}    +\sum_\xi\hbar \omega_\xi g_{\xi}Q_{m,\xi} \, .
\end{eqnarray}
Here, $\omega_\xi$ ($\xi=1,\ldots,N_{\rm vib}$) is the vibrational frequency of mode $Q_{m,\xi}$ (note the use
of dimensionless units) which is assumed to be identical for the different monomers. The same approximation is
made for the linear coupling constant, $g_\xi$, which relates to the Huang-Rhys factor as $S_{\xi} =
g_{\xi}^2/2$.

Absorption and emission spectra will be calculated in Condon-approximation for the monomeric
transition dipole matrix elements $d_m$, summed to give the aggregate dipole operator according to
\begin{equation}
  \label{eq:3}
  d=\sum_{m}d_m \ket{m}\bra{0} + {\rm h.c.} \, .
\end{equation}
\subsection{$n$-Particle Approach} 
The problem of coupled exciton-vibrational dynamics can be approached by a $n$-particle approximation scheme
\cite{Philpot}. To this end we introduce vibrational states for the different potential energy surfaces
according to $\ket{M_{m,a}}$ where it is understood that $M_{m,a}$ could be a multi-index in cases where
several vibrational modes per monomer need to be taken into account. The eigenstates $\ket{\alpha^{(1)}}$ of
$H^{(1)} (\mathbf{Q})$ can be expanded as
\begin{equation}
  \label{eq:1}
  \ket{\alpha^{(1)}} = \sum_{m,\mu}C_{m,\mu}(\alpha^{(1)}) \ket{m\mu} + \sum_{mn}\sum_{\mu \nu} C_{mn,\mu\nu}(\alpha^{(1)}) \ket{m\mu,n\nu} + \ldots
\end{equation}
with the one-particle states
\begin{equation}
  \label{eq:2}
  \ket{m\mu} = \ket{e_m}\ket{\mu=M_{m,e}} \prod_{n\ne m} \ket{g_n}\ket{0_{n,g}}
\end{equation}
and the two-particle states
\begin{equation}
  \label{eq:tpa-states}
  \ket{mn,\mu\nu} = \ket{e_m}\ket{\mu=M_{m,e}} \ket{g_n}\ket{\nu=N_{n,g}} \prod_{k\ne m,n} \ket{g_k}\ket{0_{k,g}}
\end{equation}
In principle \Eq{eq:1} will contain further terms, but for the present case of a molecular heterodimer, the
two-particle ansatz is already exact. The restriction to the first term is the so-called OPA.

Using the eigenstates, Eq.~(\ref{eq:1}), the absorption spectrum will be calculated in the zero temperature
limit as  ($I_{0}$ normalisation constant)
\begin{equation}
  \label{eq:abs_SOS}
  I(\omega)=I_0 \omega \sum_{\alpha^{(1)}} \frac{\Gamma |\langle
    \alpha^{(1)}|d|\alpha^{(0)}=0\rangle|^2}{(\omega-\omega_{\alpha^{(1)} 0})^2+\Gamma^2}\, ,
\end{equation}
where $\Gamma$ is a parameter mimicking the finite line width (dephasing time) of the real
system, $|\alpha^{(0)}\rangle$ denotes the eigenstates of $H^{(0)}$ with  $|\alpha^{(0)}=0 \rangle$ being
the overal ground state (see below), and $\omega_{\alpha^{(1)} \alpha^{(0)}}$ is the transition frequency.
The emission spectrum is calculated as
\begin{equation}
  \label{eq:em_SOS}
  F(\omega) = F_0 \sum_{\alpha^{(1)},\alpha^{(0)}}P(E_{\alpha^{(1)}}) \frac{\Gamma |\langle
    \alpha^{(1)}|d|\alpha^{(0)}\rangle|^2}{(\omega-\omega_{\alpha^{(1)}\alpha^{(0)}})^2+\Gamma^2} \, ,
\end{equation}
where $P(E_{\alpha^{(1)}})$ is the Boltzmann population of an one-exciton-vibrational state with energy $E_{\alpha^{(1)}}$ and $F_{0}$ is a normalisation constant.
\subsection{MCTDH Approach}
In principle the $n$-particle approach provides a systematic route to the exact eigenstates of the one-exciton
Hamiltonian. However, with increasing aggregate size and number of vibrational coordinates one will soon face
the dimensionality bottleneck and the problem will become numerically intractable. An efficient alternative
approach is the MCTDH method, which rests on the expansion of the time-dependent state vector into a basis of
time-dependent Hartree products that are composed of single particle functions (SPFs) \cite{beck00:1,
  meyer90:73}. Applications to exciton dynamics and spectroscopy have been given in
Refs. \cite{seibt09:13475,Ambrosek:2012ek}.

First, the state vector is expanded in terms of the one-exciton basis, i.e
\begin{eqnarray}
  |\Psi({\bf Q};t) \rangle=\sum_{m} \chi_{m}({\bf Q};t) \, |m\rangle	\, .
\end{eqnarray}
In a next step the nuclear wave function is written in MCTDH form as follows
\begin{equation}
  \label{eq:psiMCTDH}
  \chi_m(\mathbf{ Q},t) = \sum_{j_1 \ldots j_D}^{{n_{j_1} \ldots n_{j_D}}}
  C^{(m)}_{j_1,\ldots,j_D}(t) \phi^{(m)}_{j_1}(Q_1;t) \ldots \phi^{(m)}_{j_D}(Q_{D};t) \, .
\end{equation}
Here, the $C^{(m)}_{j_1,\ldots,j_D}(t)$ are the time-dependent expansion coefficients weighting the
contributions of the different Hartree products, which are composed of SPFs, $\phi^{(m)}_{j_k}(Q_k;t)$, for
the $k$th nuclear degree of freedom in state $m$. Overall there are $D=N_{\rm agg} \times N_{\rm vib}$ nuclear
degrees of freedom.

Although it is in principle possible to obtain eigenstates with the MCTDH method, being a time-dependent
approach, it is more suited to solve the time-dependent Schr\"odinger equation. Therefore we will use it to
calculate the absorption spectrum according to the time-dependent reformulation of Eq.~(\ref{eq:abs_SOS}) \cite{MayKuhnBook}
\begin{equation}
  \label{eq:abs}
  I(\omega)=I_0 \omega \, {\rm Re} \int_0^\infty dt\, e^{i\omega t-t/\tau} \langle \Psi_0 |d e^{-i H_{\rm agg} (\mathbf{Q}) t/\hbar} d | \Psi_0 \rangle \, ,
\end{equation}
where $d$ is dipole operator, Eq.~(\ref{eq:3}) and $\ket{\Psi_0}$ is the ground state wave function. Since
there is no coupling in $H^{(0)}$, it is given as a Hartree product, i.e.
\begin{equation}
  \label{eq:4}
  \ket{\Psi_0} =  \phi^{(0)}_0(Q_1) \ldots \phi^{(0)}_0(Q_{D}) \ket{0}
\end{equation}
with $\phi^{(0)}_0(Q_j)$ being the respective ground state wave function for the $j$th mode.  
All wave packet
propagation simulations have been performed using the Heidelberg program package \cite{mctdh_package}. The
MCTDH dimer setup, which has been applied here, includes three electronic states (one ground state and two
singly excited states). Since the ground state has been used for the preparation of the initial wave packet
only, one SPF per mode has been sufficient to describe it properly. The actual propagation of the wave packet
takes place on the two excited states; here four SPFs were necessary to obtain converged results. As a
primitive basis we have used a harmonic oscillator (20 points) discrete variable representation within an
interval of [-3.5:3.5] for all modes. For all calculations the multi-set method was used.
\section{Results and Discussion} 
\label{sec:results}
In the following application the case of two sites (vibronic heterodimer, $m=(D,A)$) will be considered. The
parameters are chosen to mimic the situation in the FMO complex. A number of studies have proposed electronic
Hamiltonian of the complex, based on modelling of a set of spectroscopic observables. For a review see
\cite{Milder2010}. There is a general agreement that the BChl 3 and 4 have the lowest site energies. The
transition energy difference is proposed to be from about 110 to 180 \cm. Excitonic coupling between these two
molecules is negative with values from about -50 to -75 \cm \cite{Adolphs_Renger2006}. Fluorescence line
narrowing spectroscopy has revealed that at around 180 \cm there are a few vibrational modes in the FMO lowest
energy BChl with a total Huang-Rhys factor 0.03 \cite{Wendling2000}. We point out that the analogous
experiments with BChl in triethylamine give a somewhat higher value of 0.04 \cite{Freiberg2011}.  Having these
parameters in mind we formulate our vibronic dimer as follows. The properties of the eigenstates and spectra
of the model will be scrutinised in dependence on the Coulomb coupling $J \equiv J_{\rm DA}$ and the
heterogeneity, $\Delta E \equiv E_{\rm D}-E_{\rm A}$. For reference we will use $\Delta E=180$ \cm and $J_{\rm
  DA}/hc=-90$ \cm. First, each site is coupled to one vibrational mode with parameters $\omega/2\pi c=180$ \cm
and $S=0.05$. In the following we will start with a discussion of the differences between exact and OPA in the
absorption spectrum for transitions to the one-exciton manifold. The focus will be on the dependence on
detuning between the local excitation energies and the Coulomb coupling.  In order to study multi-mode effects
a second model including additional vibrational modes with varying frequency and Huang-Rhys factor will be
considered in an up to three-mode model ($\omega/2\pi c=365$ \cm, $S=0.019$ and $\omega/2\pi c=260$ \cm,
$S=0.023$). The actual frequencies of the modes were chosen as approximately 1.5 and 2 times the original
frequency $\omega/2\pi c=180$ \cm. The corresponding Huang Rhys factors were chosen to represent the total S
of the nearby modes reported in \cite{Freiberg2011}. Finally, the temperature dependence is addressed in terms
of the emission spectrum.
\subsection{Validity of OPA} 
\begin{figure}[h]
  \centering
  \includegraphics[width=1.0\textwidth]{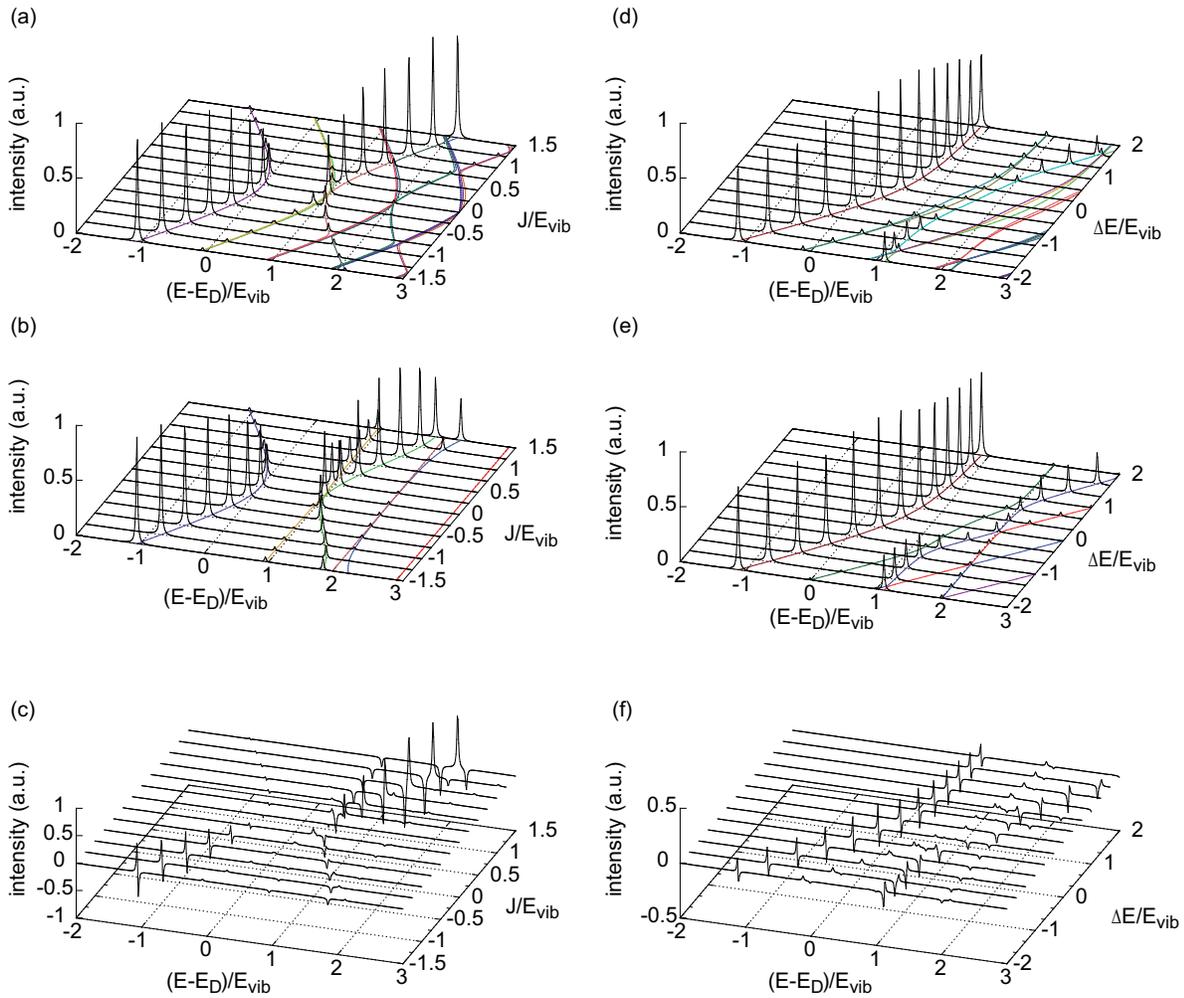}
  \caption{Absorption spectrum, Eq. (\ref{eq:abs_SOS}), as function of $J=J_{\rm DA}$ for $\Delta E/\hbar \omega=1$ (left) and as
    function of $\Delta E$ for $J_{\rm DA}/\hbar \omega=-0.5$ (right) using exact (a,d) and OPA (b,e)
    approaches. Panels (c) and (f) show the respective differences. ($E_{\rm vib}=\hbar \omega$)}
  \label{fig1}
\end{figure}

The dependence of exact and OPA absorption spectra on the Coulomb coupling $J_{\rm DA}$ ($\Delta E/\hbar
\omega=1$) and the detuning $\Delta E$ ($J_{\rm DA}/\hbar \omega = -0.5$) are compared in Fig. \ref{fig1}.
This figure also contains the respective dependencies of the bare energy levels.

First we notice that for $J_{\rm DA}=0$ the two cases show different degeneracies of energy levels due to the
restricted state space of the OPA. Looking at the dependence on $J_{\rm DA}$ upon increasing its value, energy
levels are shifted and degeneracies are partly lifted. Around $J_{\rm DA}/\hbar \omega\approx 0.85$ and
$\approx 1.35$ energy levels approach each other showing partially avoided crossings in the exact case. This
pattern is not at all reproduced by the OPA. Overall only the lowest state shows an agreement between exact
and OPA calculations. The same conclusion can be drawn from the $\Delta E$ dependence shown in
Fig. \ref{fig1}d,e.

Comparing the rather different behaviour of the spectrum of $H^{(1)}$ for exact und OPA models, the question
arises how this will reflect in absorption spectra. Here we notice from Fig. \ref{fig1} that the differences
between the exact and OPA are particularly large for the $J_{\rm DA}$-dependence. Of course, whether or not
deviations are visible depends on distribution of oscillator strength.  Therefore the positive coupling part
of the $J_{\rm DA}$ dependence is more visibly influenced. As a consequence OPA gives reasonable results as
far as the $\Delta E$-dependence is concerned since $J_{\rm DA}<0$ for the present model. Generally, there is
a trend that despite of radically different energy level structure in the two models, when it comes to the
intensities of the spectral features, the models give surprisingly similar results. If one would use broader
line shapes, the resulting spectra would not be very different for most of the used parameters. Interestingly,
it is not true that OPA necessarily yields a simpler spectrum as one would expect on the basis of the energy
level structure of the one-exciton Hamiltonian.  Indeed, one can state that the exact case shows more
pronounced collective effect, i.e. oscillator strength is essentially located in a single transition.
\subsection{Multi-mode Effects} 
\label{sec:multi-mode-effects}
\begin{figure}[h]
  \label{fig:abs_mctdh}
  \centering
  \includegraphics[width=1\textwidth]{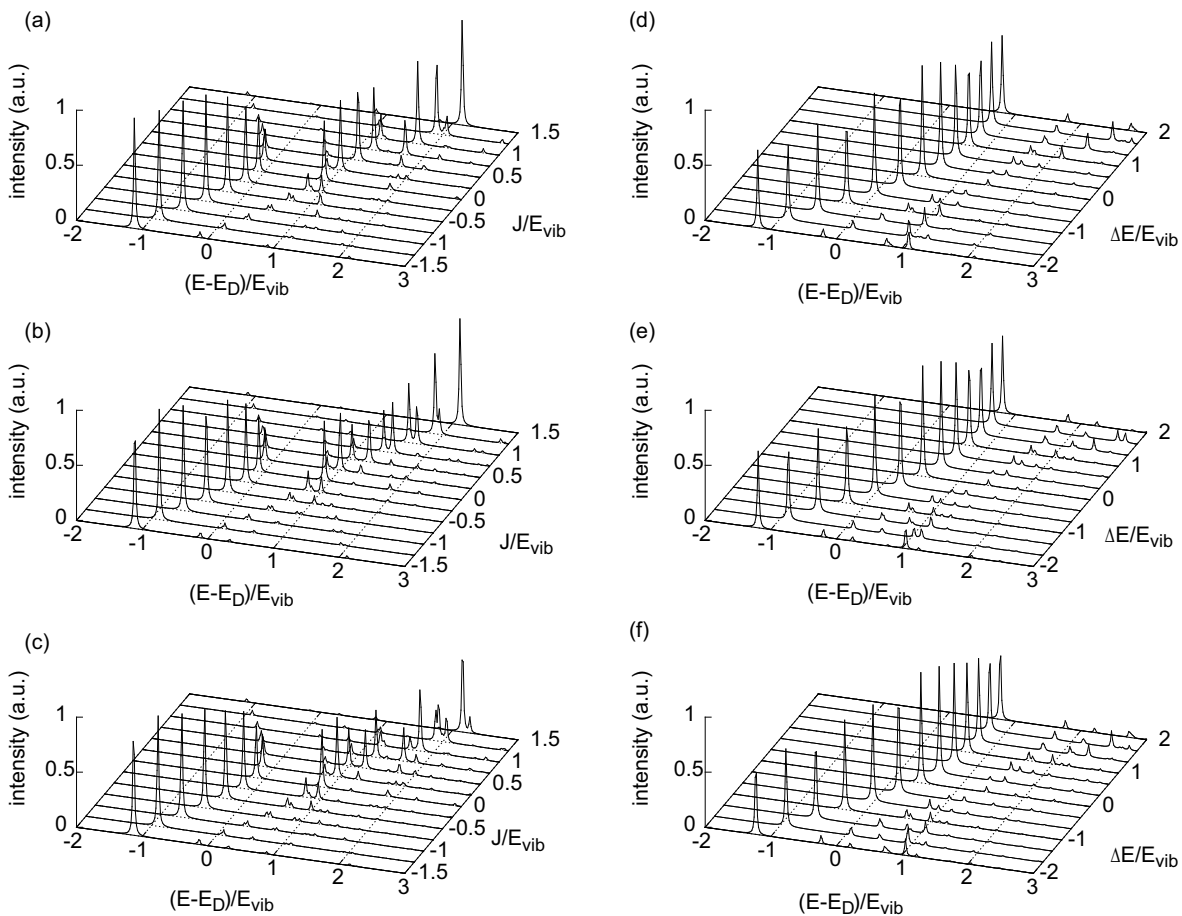}
  \caption{Absorption spectrum, Eq. (\ref{eq:abs}), as function of $J=J_{\rm DA}$ for $\Delta E/\hbar \omega=1$ (left) and as
    function of $\Delta E$ for $J_{\rm DA}/\hbar \omega=-0.5$ (right) for two different two-mode models and a
    three-mode model. All results contain the mode at 180 \cm used before. Additional modes: (a,d)
    $\omega/2\pi c=365$ \cm and $S=0.019$, (b,e) $\omega/2\pi c=260$ \cm and $S=0.023$, (c,f) both modes of
    panels (a-d). ($E_{\rm vib}=\hbar \omega$, 180 \cm mode)}
\end{figure}
In the following we discuss the effect of multiple modes, coupling to the electronic transitions. Results of
MCTDH simulations of absorption spectra are shown in Fig. 2. Overall one can state that since for large
negative $J_{\rm DA}$ oscillator strength is concentrated in the lowest state which has dominantly electronic
character, a pronounced effect of further modes is seen for positive couplings only. For the case of FMO this
leads to rather similar spectra as a function of the detuning $\Delta E$ as can be seen from the right column
of Fig. 2. In case of positive coupling the details of spectral changes depend, of course, on the mode
parameters, but the net effect is a broadening due to the more complex structure of the exciton-vibronic
states.
\subsection{Temperature Dependence of Emission Spectra} 
\label{sec:temp-depend-absorpt}
\begin{figure}[h]
  \centering
  \includegraphics[width=\textwidth]{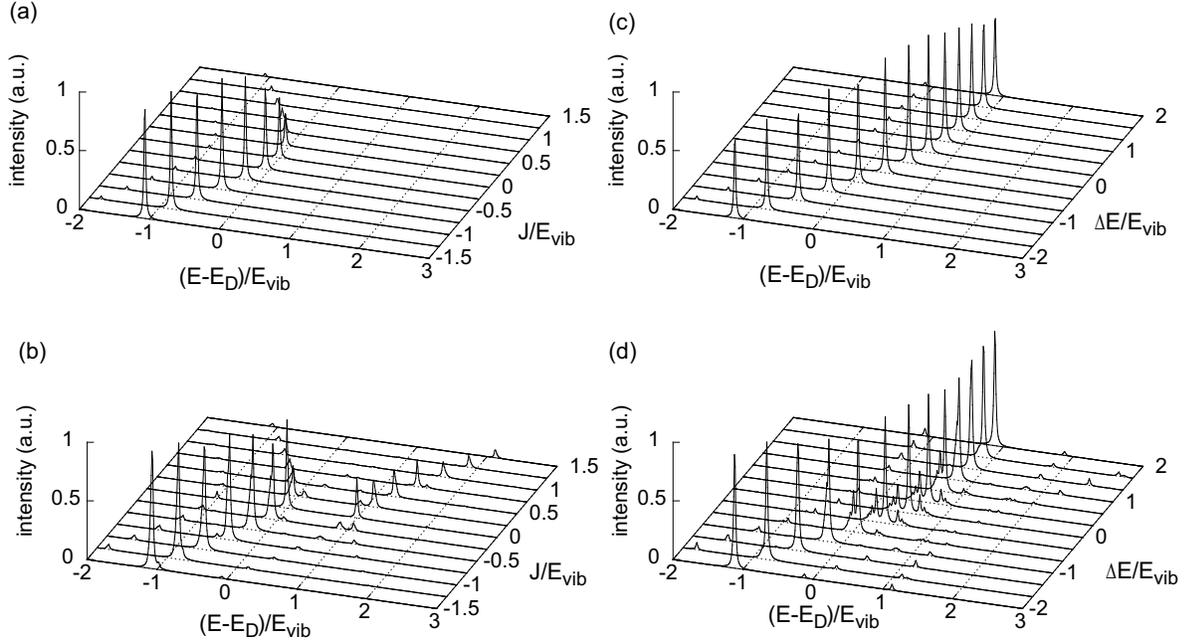}
  \caption{Emission spectra, Eq. (\ref{eq:em_SOS}), as function of $J=J_{\rm DA}$ for $\Delta E/\hbar \omega=1$ and $T=4$K (a) and 300
    K (b) and as a function of $\Delta E$ for $J_{\rm DA}/\hbar \omega=-0.5$ and $T=4$K (c) and 300 K
    (d). ($E_{\rm vib}=\hbar \omega$)}
  \label{fig:emi}
\end{figure}

Results for the emission spectrum of the one-mode model as a function of Coulomb coupling and detuning are
shown in Fig. \ref{fig:emi} for two different temperatures.  Owing to the small Huang-Rhys factor the spectra
at 4K essentially are showing the 0-0 transition and a small shoulder due to the 0-1 transition in
Fig. \ref{fig:emi}a. The intensity of these transitions gradually drops down upon increasing $J_{\rm DA}$ in
the positive domain due to H-aggregate formation. This does not occur upon increasing the detuning in panel
(c) where the spectrum becomes close to that of a monomer for large $\Delta E$. For $T=300 $K the spectrum is
much more structured as can be seen from Fig. \ref{fig:emi}b,d. The level structure with a complex intensity
pattern makes it impossible to draw \emph{a priori} conclusions on the thermal occupation. Notice that due to
the Coulomb coupling the spacing between the peaks above the 0-0 transition is not equal to a vibrational
quantum. Since the lowest energy fluorescence feature is due to the ground state vibrational structure, it is
still shifted from 0-0 by the vibrational frequency.
\subsection{Excitonic Distortions of Fluorescence Site Selection Spectra} 

\begin{figure}[h]
  \centering
  \includegraphics[width=0.75\textwidth]{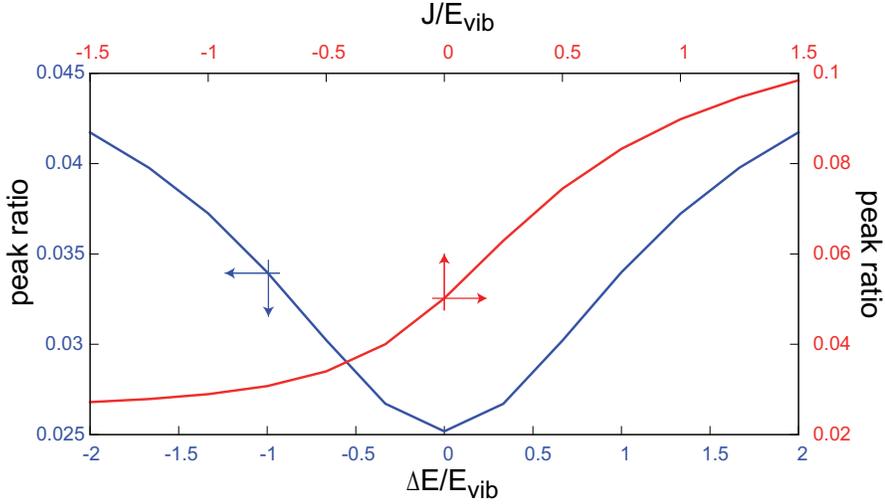}
  \caption{Peak ratios between 0-0 and 0-1 transitions at $T=4$K as a function of $J=J_{\rm DA}$ for $\Delta
    E/\hbar \omega=1$ and of $\Delta E$ for $J_{\rm DA}/\hbar \omega=-0.5$ (cf. Fig. \ref{fig:emi}). The
    arrows mark the uncoupled case (red) and the detuning corresponding to the FMO model (blue). ($E_{\rm
      vib}=\hbar \omega$)}
  \label{fig:ratio}
\end{figure}
The ratio between 0-0 and 0-1 emission intensities is investigated in more detail in Fig. \ref{fig:ratio}. For
a monomer this ratio would be equal to the Huang-Rhys factor what was used in the calculations, $S=0.05$ in
the present case. This value is observed only for $J_{\rm DA}=0$ (marked as a crossing of red arrows in Fig. \ref{fig:ratio})  and for a finite $J_{\rm DA}$ in the limit of
large detunings (not shown). At this point we should recall that Huang-Rhys factors and mode frequencies are
usually obtained from low-temperature site-selected fluorescence spectra. Clearly, the outcome of such
experiment would be influenced by the excitonic coupling. The case of FMO would approximately correspond to
the blue curve of the Fig. 4 at the relative detuning -1 marked as a crossing of blue arrows.  Our calculations give for that point the "effective
observable" of about S=0.034. We point out that the Huang-Rhys factor obtained from the FMO experiment is
smaller than the $S$ reported for BChl in solution. Our calculations give a straightforward explanation of
this discrepancy. With the help of our results illustrated in Fig. \ref{fig:ratio} the Huang-Rhys factors
obtained from fluorescence site selection spectroscopy of light-harvesting complexes and other molecular
aggregates can be corrected. Obviously the correction can easily be as large as 100\%. In case of the FMO in 180 \cm region the observed S should be multiplied by a factor of 1.5. The correction factor becomes smaller for the higher frequency modes. Theoretical studies where the spectral density is extracted from atomic simulations and compared to the fluorescence spectra for benchmarking \cite{Ulrich, Shim2012}, need to consider these correction factors. 

Finally, we note that
the measured $S$ values will be reduced in case of negative $J_{\rm DA}$ values (J-aggregate) and increased
for the opposite sign (H-aggregate).

\section{Summary}
\label{sec:sum}

We have carried out a thorough comparison of OPA and exact calculations of vibronic excitons in a dimer
model. We found that the general behaviour of the calculated absorption spectra are surprisingly similar
despite of the radically different energy level structure in the two models. Multi-mode effects have been
discussed in dependence on the mode parameters. In the case of negative coupling, which is relevant for the
FMO complex, most oscillator strength is concentrated in a transition to a state of essentially electronic
character such that no pronounced effect on the spectrum arises. Calculations of the fluorescence spectra show
that the Huang Rhys factors obtained from fluorescence spectroscopy of light-harvesting complexes with
significant excitonic couplings, need to be corrected. Depending on the excitonic coupling strength and
detuning, the correction can be as large as 100\%. At the same time, the observed frequencies are not affected
if measured at low temperature where the transitions from higher energy levels are not giving any significant
contribution.

\section*{Acknowledgments}
The authors acknowledge support from KAW, Swedish Energy Agency, Swedish Research Council, the Deutsche
Forschungsgemeinschaft (Sfb652) and EU Erasmus program.
\section*{References}

\end{document}